\documentclass{ws-procs9x6}

\begin{document}
\setcounter{page}{103}
\title{Justification for the \\ composite fermion picture}

\author{A. W\'OJS,$^{1,2}$ J.~J. QUINN,$^1$ AND L. JACAK$^2$}
\address{
   $^1$University of Tennessee, Knoxville, Tennessee 37996, USA\\
   $^2$Wroclaw University of Technology, Wroclaw 50-370, Poland\\
   E-mail: wojs@if.pwr.wroc.pl}


\maketitle

\abstracts{
The mean field (MF) composite Fermion (CF) picture successfully 
predicts the low-lying bands of states of fractional quantum Hall 
systems.
This success cannot be attributed to the originally proposed 
cancellation between Coulomb and Chern--Simons interactions beyond 
the mean field and solely depends on the short range of the repulsive
Coulomb pseudopotential in the lowest Landau level (LL).
The class of pseudopotentials is defined for which the MFCF picture 
can be applied.
The success or failure of the MFCF picture in various systems 
(electrons in the lowest and excited LL's, Laughlin quasiparticles) 
is explained.}

\section{Introduction}

The quantum Hall effect (QHE)\cite{klitzing,tsui} is the quantization 
of Hall conductance of a two-dimensional electron gas (2DEG) in high 
magnetic fields that occurs at certain fillings of the macroscopically 
degenerate single-electron Landau levels (LL's).
The LL filling is defined by the filling factor $\nu$ equal to the
number of electrons divided by the LL degeneracy, which is proportional
to the magnetic field and the physical area occupied by the 2DEG.
The series of values of $\nu$ at which the QHE is observed contains
small integers and simple, almost exclusively odd-denominator fractions
such as $\nu={1\over3}$, ${2\over5}$, etc.
The series is universal for all samples, which means that the occurrence 
of QHE at a particular value of $\nu$ depends on the quality of the 
sample, temperature, or other similar conditions, but not on the material
parameters such as lattice composition.
Moreover, the observed values of $\nu$ are ``exact'' in a sense that 
poor sample quality (e.g., lattice imperfections) can destroy 
QHE at some or all of the values of $\nu$, but cannot shift these 
values.

The quantization of Hall conductance is always accompanied by a rapid
drop of longitudinal conductance, and both effects signal the appearance 
of (incompressible) nondegenerate many-body ground states (GS's) in 
the spectrum of the 2DEG, separated from the continuum of excited states 
by a finite gap.
At integer $\nu=1$, 2 \dots\ (IQHE), the origin of incompressibility 
is the single particle cyclotron gap between the LL's.
On the other hand, at fractional $\nu=1/3$, 1/5, 2/5 \dots\ (FQHE) 
electrons partially fill a degenerate (lowest) LL and the formation 
of incompressible GS's is a complicated many-body phenomenon.
The gap and resulting incompressibility are entirely due to 
electron-electron (Coulomb) interactions and reveal the 
unique properties of this interaction within the lowest 
LL.\cite{laughlin1,prange}

In this note we shall concentrate on the latter effect.
We will apply the pseudopotential formalism\cite{haldane1,parent} 
to the FQH systems, and show that the form of the pseudopotential 
$V(L')$ [pair energy vs.\ pair angular momentum] in the lowest LL 
rather than of the interaction potential $V(r)$, is responsible 
for incompressibility of the FQH states.
The idea of fractional parentage\cite{shalit} will be used to 
characterize many-body states by the ability of electrons to avoid 
pair states with largest repulsion.
The condition on the form of $V(L')$ necessary for the occurrence 
of FQH states will be given, which defines the short-range repulsive 
(SRR) pseudopotentials to which MFCF picture can be applied.
As an example, we explain the success or failure of MFCF predictions 
for the electrons in the lowest and excited LL's and for Laughlin 
quasiparticles (QP's) in hierarchy picture of FQH 
states.\cite{haldane2,sitko}

\section{Theoretical concepts for FQHE}

\subsection{Single-electron states and Laughlin wavefunction}

The Hamiltonian for an electron confined to the $x$--$y$ plane in 
the presence of a perpendicular magnetic field $B$ is
\begin{equation}
   H_0={1\over2\mu}\left(\vec{p}+{e\over c}\vec{A}\right)^2.
\end{equation}
Here $\mu$ is the effective mass, $\vec{p}=(p_x,p_y,0)$ is the momentum 
operator and $\vec{A}(x,y)$ is the vector potential (whose curl gives $B$).
For the ``symmetric gauge,'' $\vec{A}={1\over2}B(-y,x,0)$, the single
particle eigenfunctions\cite{gasiorowicz} are of the form $\psi_{nm}
(r,\theta)=e^{-im\theta}u_{nm}(r)$, and the eigenvalues are given by
\begin{equation}
   E_{nm}={1\over2}\hbar\omega_c(2n+1+|m|-m).
\end{equation}
In these equations, $n=0$, 1, 2, \dots, and $m=0$, $\pm1$,  $\pm2$, 
\dots.
The lowest energy states (lowest Landau level) have $n=0$, $m=0$, 
1, 2, \dots, and energy $E_{0m}=\hbar\omega_c/2$.
It is convenient to introduce a complex coordinate $z=re^{-i\theta}
=x-iy$, and to write the lowest Landau level wavefunctions as 
$\psi_{0,-m}=N_mz^m\exp(-|z|^2/4)$, where $N_m$ is a normalization 
constant, and $m$ can take on any non-negative integral value.
In this expression we have used the magnetic length $\lambda=(\hbar 
c/eB)^{1/2}$ as the unit of length.
The function $|\psi_{0,-m}|^2$ has its maximum at a radius 
$r_m$ which is proportional to $m^{1/2}$.
All single particle states from a given Landau level are 
degenerate, and separated in energy from neighboring levels by
$\hbar\omega_c$.

If $m$ is restricted to being less than some maximum value, $N_L$,
chosen so that the system has a ``finite radial range,'' then the
alowed $m$ values are 0, 1, 2, \dots, $N_L-1$.
The value of $N_L$ is equal to the flux through the sample, 
$B\cdot A$ (where $A$ is the area), divided by the quantum of flux
$\phi_0=hc/e$.
The filling factor $\nu$ is defined as the ratio of the number of 
electrons, $N$, to $N_L$.
An infinitesimal decrease in the area $A$ when $\nu$ has an integral 
value requires promotion of an electron across the gap $\hbar\omega_c$
to the first unoccupied level, making the system incompressible.

In order to construct a many electron wavefunction corresponding to
filling factor $\nu=1$, the product function which places one electron
in each of the $N_L$ orbitals $\psi_{0,-m}$ ($m=0$, 1, \dots, $N_L-1$)
must be antisymmetrized.
This can be done with the aid of a Slater determinant
\begin{equation}
\label{vandemonde}
   \Psi(z_1,z_2,\dots,z_N)\propto\left| 
   \matrix{ 1 & 1 & \dots & 1 \cr
            z_1 & z_2 & \dots & z_N \cr
            z_1^2 & z_2^2 & \dots & z_N^2 \cr
            \vdots & \vdots & & \vdots \cr
            z_1^{N_L-1} & z_2^{N_L-1} & \dots & z_N^{N_L-1} }
   \right| e^{-\sum_k|z_k|^2/4}.
\end{equation}
The determinant in Eq.~(\ref{vandemonde}) is the well-known Vandemonde 
determinant.
It is not difficult to show that it is equal to $\prod_{i<j}(z_i-z_j)$.
Of course, $N_L=N$ (since each of the $N_L$ orbitals is occupied by 
one electron) and $\nu=1$.

Laughlin noticed that if the factor $(z_i-z_j)$ of Vandemonde 
determinant was replaced by $(z_i-z_j)^m$, where $m$ was an odd
integer, the wavefunction 
\begin{equation}
   \Psi_m(z_1,z_2,\dots,z_N)
   \propto\prod_{i<j}(z_i-z_j)^m
   e^{-\sum_k|z_k|^2/4}
\end{equation}
would be antisymmetric, keep the electrons further apart (and therefore
reduce repulsion), and correspond to a filling factor $\nu=m^{-1}$.
This results because the highest power of the orbital index entering
$\Psi_m$ is $N_L-1=m(N-1)$ giving $\nu=N/N_L=m^{-1}$ in the limit of 
large systems.
The additional factor $\prod_{i<j}(z_i-z_j)^{m-1}$ multiplying
$\Psi_{m=1}(z_1,z_2,\dots,z_N)$ is the Jastrow factor which accounts 
for correlations.

\subsection{Laughlin quasiparticles and Haldane hierarchy}

The elementary charged excitations of the Laughlin $\nu=(2p+1)^{-1}$
GS are Laughlin quasiparticles (QP's) corresponding to a vortex (for 
the quasihole, QH) or anti-vortex (for the quasielectron, QE) at an 
arbitrary point $z_0$, and described by the following wavefunctions,
\begin{equation}
   \Psi_{\rm QH}\propto(z-z_0)\Psi_m
	\;\;\mbox{and}\;\;
   \Psi_{\rm QE}\propto{\partial\over\partial(z-z_0)}\Psi_m.
\end{equation}
Laughlin QP's carry fractional electric charge $\pm e/(2p+1)$ 
and obey fractional statistics (although in some situations
they can be conveniently treated as either fermions or bosons 
thanks to a statistics transformation valid in two dimensions).
Being charged particles (of the finite size of the order of the
magnetic length $\lambda$) moving in the magnetic field, Laughlin
QP form (quasi)LL's similar to those of electrons, except for 
the $m$-times lower degeneracy due to theor reduced charge. 
They are also naturally expected to interact with one another
via normal, charge-charge Coulomb forces.

As an extension of Laughlin's idea, Haldane,\cite{haldane2} 
and others\cite{sitko,laughlin2} proposed that in analogy to 
electrons, Laughlin QP's must form Laughlin-like incompressible 
states of their own.
According to this idea, each Laughlin state of electrons would stand
atop entire family of so-called ``daughter'' Laughlin state of its
QE's and QH's.
And on the following level, each daughter state would have its own
family of daughter states, and so on.
This construction results in entire family of incompressible states
that corresponds to many more fractional filling factors $\nu$ at
which incompressibility and in result also the FQHE are expected
in the underlying 2DEG.
For example, the $\nu={2\over5}$ state can be interpreted as 
$\nu={1\over3}$ QE daugter state of the parent $\nu={1\over3}$ 
electron state.
In fact, all odd-denominator fractions $\nu=p/q$ can be generated 
in this way, which brings us to the major problem of the (original) 
concept of hierarchy.
On one hand, the hierarchy predicts too many fractions (only a finite
number of fractions are observed experimentally, and it seems evident
that FQHE will not occur at most of the other predicted fractions 
regarless of the experimental conditions\cite{tsui,willet,mallet}).
On the other hand, the hierarchy model gives no apparent connection 
between the stability of a given state and its position in the hierarchy
(explanation of some of the easily experimentally observed FQH states
requires introducing many generations of QP's).
As we shall explain later, these problems of the hierarchy model
resulted from an erroneous assumption that, being charged particles,
Laughlin QP's will form Laughlin states at each Laughlin filling
factor, $\nu=(2p+1)^{-1}$.
With the knowledge of the form of QP--QP interactions, one can
eliminate ``false'' daughter states from the hierarchy and reach
an agreement with the experimental observation.
This makes the Laughlin--Haldane theory the only microscopic
theory of the FQHE.

\subsection{Jain composite Fermion model}

Independently of the Laughlin--Haldane model, from the similar 
energy spectra of the FQH and IQH systems one can expect that 
some kind of effective, charged particle-like excitations may 
form in the interacting 2DEG.
These excitations would be the relevant charge carriers near 
$\nu={1\over3}$ and they would fill exactly their quasi-LL's at 
precisely this value, giving rise to incompressibility.
This idea leads to the composite Fermion (CF) picture.\cite{jain,lopez}

In the mean field (MF) CF picture, in a 2DEG of density $n$ at a strong 
magnetic field $B$, each electron is assumed to bind an even number $2p$ 
of magnetic flux quanta $\phi_0=hc/e$ (in form of an infinitely thin 
flux tube) forming a CF.
Because of the Pauli exclusion principle, the magnetic field confined
into a flux tube within one CF has no effect on the motion of other CF's, 
and the average effective magnetic field $B^*$ seen by CF's is reduced, 
$B^*=B-2p\phi_0n$.
Because $B^*\nu^*=B\nu=n\phi_0$, the relation between the electron and 
CF filling factors is
\begin{equation}
   (\nu^*)^{-1}=\nu^{-1}-2p.
\end{equation}
Since the low band of energy levels of the original (interacting) 2DEG 
has similar structure to that of the noninteracting CF's in a uniform 
effective field $B^*$, it was proposed\cite{jain} that the Coulomb
charge-charge and Chern--Simons (CS) charge-flux interactions beyond 
the MF largely cancel one another, and the original strongly interacting 
system of electrons is converted into one of weakly interacting CF's.
Consequently, the FQHE of electrons was interpreted as the IQHE of CF's.

Although the MFCF picture correctly predicts the structure of low-energy
spectra of FQH systems, the energy scale it uses (the CF cyclotron energy
$\hbar\omega_c^*$) is totally irrelevant.
Moreover, since the characteristic energies of CS ($\hbar\omega_c^*
\propto B$) and Coulomb ($e^2/\lambda\propto\sqrt{B}$, where $\lambda$ 
is the magnetic length) interactions between fluctuations beyond MF scale
differently with the magnetic field, the reason for its success cannot 
be found in originally suggested cancellation between those interactions.
Since the MFCF picture is commonly used to interpret various numerical 
and experimental results, it is important to understand why and under 
what conditions it is correct.

\section{Numerical exact diagonalization studies}

Because of the LL degeneracy, the electron-electron interaction in the 
FQH states cannot be treated perturbatively, and the exact (numerical) 
diagonalization techniques have been commonly used in their study.
In order to model an infinite 2DEG by a finite (small) system that can 
be handled numerically, it is very convenient to confine $N$ electrons
to a surface of a (Haldane) sphere of radius $R$, with the normal magnetic 
field $B$ produced by a magnetic monopole of integer strength $2S$ (total 
flux of $4\pi BR^2=2S\phi_0$) in the center.\cite{haldane2}
The obvious advantages of such geometry is the absence of an edge and 
preserving full 2D symmetry of a 2DEG (good quantum numbers are the total 
angular momentum $L$ and its projection $M$).
The numerical experiments in this geometry have shown that even relatively 
small systems that can be solved exactly on a computer behave in many 
ways like an infinite 2DEG, and a number of parameters of a 2DEG 
(e.g. excitation energies) can be obtained from such small 
scale calculations.

The single particle states on a Haldane sphere (monopole harmonics) are 
labeled by angular momentum $l$ and its projection $m$.\cite{wu}
The energies, $\varepsilon_l=\hbar\omega_c[l(l+1)-S^2]/2S$, fall into
degenerate shells and the $n$th shell ($n=l-|S|=0$, 1, \dots) corresponds 
to the $n$th LL.
For the FQH states at filling factor $\nu<1$, only the lowest, spin 
polarized LL need be considered.

\begin{figure}[t]
\epsfxsize=3.5in
\centerline{\epsfbox{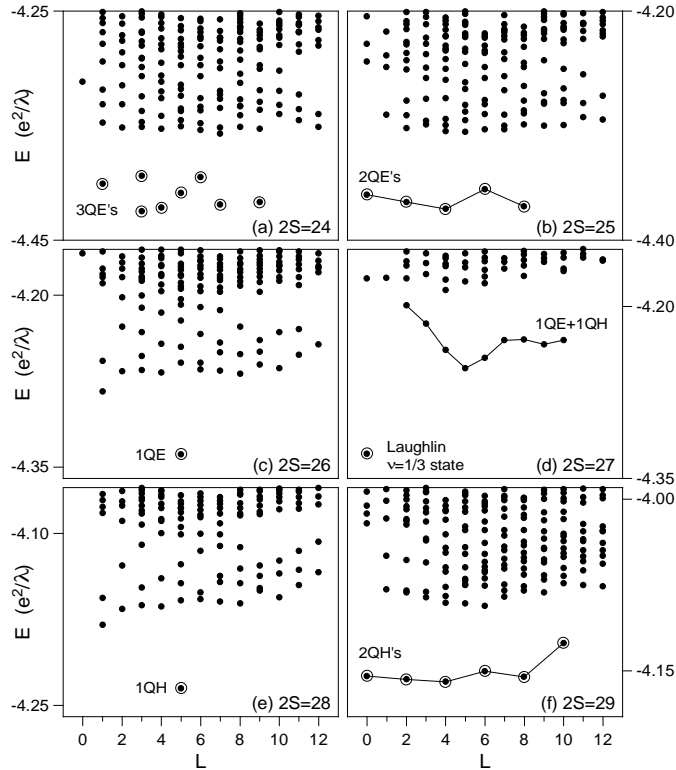}}
\caption{
   Energy spectra of ten electrons in the lowest LL at the monopole 
   strength $2S$ between 24 and 29. 
   Open circles mark lowest energy bands with fewest CF QP's.}
\label{fig1}
\end{figure}

The object of numerical studies is to diagonalize the electron-electron
interaction Hamiltonian $H$ in the space of degenerate antisymmetric $N$ 
electron states of a given (lowest) LL.
Although matrix $H$ is easily block diagonalized into blocks with specified 
$M$, the exact diagonalization becomes difficult (matrix dimension over 
$10^6$) for $N>10$ and $2S>27$ ($\nu=1/3$).\cite{parent}
Typical results for ten electrons at filling factors near $\nu=1/3$ are 
presented in Fig.~\ref{fig1}.
Energy $E$, plotted as a function of $L$ in the magnetic units, includes 
shift $-(Ne)^2/2R$ due to charge compensating background.
There is always one or more $L$ multiplets (marked with open circles) 
forming a low-energy band separated from the continuum by a gap.
If the lowest band consists of a single $L=0$ GS (Fig.~\ref{fig1}d),
it is expected to be incompressible in the thermodynamic limit (for 
$N\rightarrow\infty$ at the same $\nu$) and an infinite 2DEG at this 
filling factor is expected to exhibit the FQHE.

The MFCF interpretation of the spectra in Fig.~\ref{fig1} is the following.
The effective magnetic monopole strength seen by CF's is\cite{jain,parent}
\begin{equation}
   2S^*=2S-2p(N-1),
\end{equation}
and the angular momenta of lowest CF shells (CF LL's) are 
$l^*_n=|S^*|+n$.\cite{chen}
At $2S=27$, $l_0^*=9/2$ and ten CF's fill completely the lowest CF shell 
($L=0$ and $\nu^*=1$).
The excitations of the $\nu^*=1$ CF GS involve an excitation of at least
one CF to a higher CF LL, and thus (if the CF-CF interaction is weak on 
the scale of $\hbar\omega_c^*$) the $\nu^*=1$ GS is incompressible and 
so is Laughlin\cite{laughlin1} $\nu=1/3$ GS of underlying electrons.
The lowest lying excited states contain a pair of QP's: a quasihole (QH) 
with $l_{\rm QH}=l^*_0=9/2$ in the lowest CF LL and a quasielectron (QE) 
with $l_{\rm QE}=l^*_1=11/2$ in the first excited one.
The allowed angular momenta of such pair are $L=1$, 2, \dots, 10.
The $L=1$ state usually has high energy and the states with $L\ge2$ form 
a well defined band with a magnetoroton minimum at a finite value of $L$.
The lowest CF states at $2S=26$ and 28 contain a single QE and a single 
QH, respectively (in the $\nu^*=1$ CF state, i.e. the $\nu=1/3$ electron
state), both with $l_{\rm QP}=5$, and the excited states will contain 
additional QE-QH pairs.
At $2S=25$ and $29$ the lowest bands correspond to a pair of QP's, and 
the values of energy within those bands define the QP-QP interaction
pseudopotential $V_{\rm QP}$.
At $2S=25$ there are two QE's each with $l_{\rm QE}=9/2$ and the allowed
angular momenta (of two identical Fermions) are $L=0$, 2, 4, 6, and 8,
while at $2S=29$ there are two QH's each with $l_{\rm QH}=11/2$ and 
$L=0$, 2, 4, 6, 8, and 10.
Finally, at $2S=24$, the lowest band contains three QE's each with 
$l_{\rm QE}=4$ and $L=1$, $3^2$, 4, 5, 6, 7, and 9.

\section{Pseudopotential and fractional grandparentage}

The two body interaction Hamiltonian $H$ can be expressed as
\begin{equation}
   \hat{H}=\sum_{i<j} \sum_{L'} V(L')\;\hat{\bf P}_{ij}(L'),
\end{equation}
where $V(L')$ is the interaction pseudopotential\cite{haldane1} and 
$\hat{\bf P}_{ij}(L')$ projects onto the subspace with angular momentum 
of pair $ij$ equal to $L'$.
For electrons confined to a LL, $L'$ measures the average squared 
distance $d^2$,\cite{parent}
\begin{equation}
   {\hat{d}^2\over R^2}=
   2+{S^2\over l(l+1)}\left(2-{\hat{L}'^2\over l(l+1)}\right),
\end{equation}
and larger $L'$ corresponds to smaller separation.
Due to the confinement of electrons to one (lowest) LL, interaction 
potential $V(r)$ enters Hamiltonian $H$ only through a small number 
of pseudopotential parameters $V(2l-{\bf R})$, where ${\bf R}$, 
relative pair angular momentum, is an odd integer.

\begin{figure}[t]
\epsfxsize=3.5in
\centerline{\epsfbox{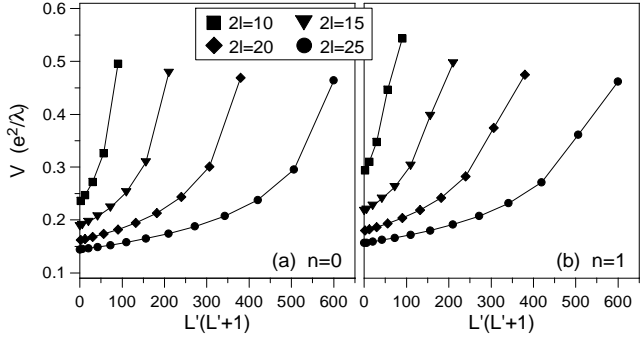}}
\caption{
   Pseudopotentials $V$ of the Coulomb interaction in the lowest (a), 
   and first excited LL (b) as a function of squared pair angular 
   momentum $L'(L'+1)$.
   Different symbols mark data for different $S=l+n$.}
\label{fig2}
\end{figure}

In Fig.~\ref{fig2} we compare Coulomb pseudopotentials $V(L')$ calculated 
for a pair of electrons on the Haldane sphere each with $l=5$, 15/2, 10, 
and 25/2, in the lowest and first excited LL.
For the reason that will become clear later, $V(L')$ is plotted as 
a function of $L'(L'+1)$.
All pseudopotentials in Fig.~\ref{fig2} increase with increasing $L'$.
If $V(L')$ increased very quickly with increasing $L'$ (we define ideal 
short-range repulsion, SRR, as: 
$dV_{\rm SR}/dL'\gg0$ and $d^2V_{\rm SR}/dL'^2\gg0$), 
the low-lying many-body states would be the ones maximally avoiding 
pair states with largest $L'$.\cite{haldane1,parent}
At filling factor $\nu=1/m$ ($m$ is odd) the many-body Hilbert space 
contains exactly one multiplet in which all pairs completely avoid 
states with $L'>2l-m$.
This multiplet is the $L=0$ incompressible Laughlin state\cite{laughlin1} 
and it is an exact GS of $V_{\rm SRR}$.

The ability of electrons in a given many-body state to avoid strongly 
repulsive pair states can be conveniently described using the idea of 
fractional parentage.\cite{parent,shalit}
An antisymmetric state $\left|l^N,L\alpha\right>$ of $N$ electrons each 
with angular momentum $l$ that are combined to give total angular 
momentum $L$ can be written as
\begin{equation}
   \left|l^N,L\alpha\right>=
   \sum_{L'}\sum_{L''\alpha''} 
   G_{L\alpha}^{L''\alpha''}(L')
   \left|l^2,L';l^{N-2},L''\alpha'';L\right>.
\end{equation}
Here, $\left|l^2,L';l^{N-2},L''\alpha'';L\right>$ denote product states 
in which $l_1=l_2=l$ are added to obtain $L'$, $l_3=l_4=\dots=l_N=l$ are 
added to obtain $L''$ (different $L''$ multiplets are distinguished by 
a label $\alpha''$), and finally $L'$ is added to $L''$ to obtain $L$.
The state $\left|l^N,L\alpha\right>$ is totally antisymmetric, and states 
$\left|l^2,L';l^{N-2},L''\alpha'';L\right>$ are antisymmetric under 
interchange of particles 1 and 2, and under interchange of any pair of 
particles 3, 4, \dots\ $N$.
The factor $G_{L\alpha}^{L''\alpha''}(L')$ is called the coefficient of 
fractional grandparentage (CFGP).
The two-body interaction matrix element is expressed as
\begin{equation}
   \left<l^N,L\alpha\right|V\left|l^N,L\beta\right>=
   {N(N-1)\over2}\!\!
	\sum_{L';L''\alpha''} 
   G_{L\alpha}^{L''\alpha''}(L') \, G_{L\beta}^{L''\alpha''}(L') \, V(L'),
\end{equation}
and expectation value of energy is
\begin{equation}
   E_\alpha(L)={N(N-1)\over2}
   \sum_{L'} {\bf G}_{L\alpha}(L') \, V(L'),
\end{equation}
where the coefficient
\begin{equation}
   {\bf G}_{L\alpha}(L')=
   \sum_{L''\alpha''}\left|G_{L\alpha}^{L''\alpha''}(L')\right|^2
\end{equation}
gives the probability that pair $ij$ is in the state with $L'$.

\section{Energy spectra of short-range repulsive pseudopotentials}

The very good description of actual GS's of a 2DEG at fillings $\nu=1/m$ 
by the Laughlin wavefunction (overlaps typically larger that 0.99) and 
the success of the MFCF picture at $\nu<1$ both rely on the fact that 
pseudopotential of Coulomb repulsion in the lowest LL falls into the 
same class of SRR pseudopotentials as $V_{\rm SRR}$.
Due to a huge difference between all parameters $V_{\rm SRR}(L')$, the
corresponding many-body Hamiltonian has the following hidden symmetry:
the Hilbert space ${\bf H}$ contains eigensubspaces ${\bf H}_p$ of 
states with ${\bf G}(L')=0$ for $L'>2(l-p)$, i.e. with $L'<2(l-p)$.
Hence, ${\bf H}$ splits into subspaces $\tilde{\bf H}_p={\bf H}_p
\setminus{\bf H}_{p+1}$, containing states that do not have 
grandparentage from $L'>2(l-p)$, but have some grandparentage from 
$L'=2(l-p)-1$,
\begin{equation}
   {\bf H}=\tilde{\bf H}_0\oplus\tilde{\bf H}_1\oplus
            \tilde{\bf H}_2\oplus\dots
\end{equation}
The subspace $\tilde{\bf H}_p$ is not empty (some states with $L'<2(l-p)$ 
can be constructed) at filling factors $\nu\le(2p+1)^{-1}$.
Since the energy of states from each subspace $\tilde{\bf H}_p$ is 
measured on a different scale of $V(2(l-p)-1)$, the energy spectrum 
splits into bands corresponding to those subspaces.
The energy gap between the $p$th and $(p+1)$st bands is of the order of 
$V(2(l-p)-1)-V(2(l-p-1)-1)$, and hence the largest gap is that between 
the 0th band and the 1st band, the next largest is that between the 1st 
band and 2nd band, etc.

\begin{figure}[t]
\epsfxsize=3.5in
\centerline{\epsfbox{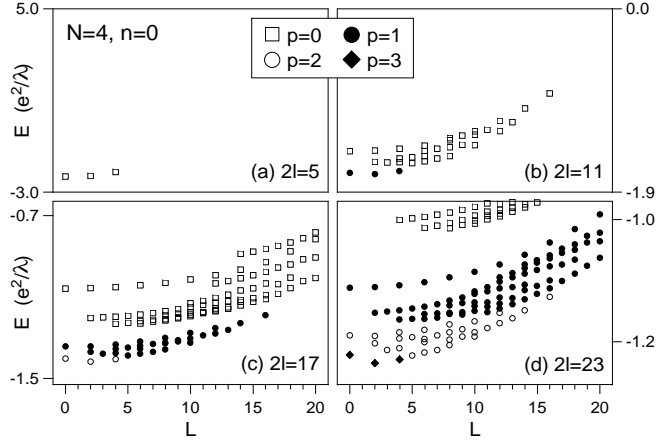}}
\caption{
   Energy spectra of four electrons in the lowest LL each with angular 
   momentum $l=5/2$ (a), $11/2$ (b), $17/2$ (c), and $23/2$ (d).
   Different subspaces ${\bf H}_p$ are marked with squares ($p=0$),
   full ($p=1$) and open circles ($p=2$), and diamonds ($p=3$).}
\label{fig3}
\end{figure}

Fig.~\ref{fig3} demonstrates on the example of four electrons to what 
extent this hidden symmetry holds for the Coulomb pseudopotential in 
the lowest LL.
The subspaces ${\bf H}_p$ are identified by calculating CFGP's of 
all states.
They are not exact eigenspaces of the Coulomb interaction, but the mixing 
between different ${\bf H}_p$ is weak and the coefficients ${\bf G}(L')$ 
for $L'>2(l-p)$ are indeed much smaller in states marked with a given $p$ 
than in all other states.
For example, for $2l=11$, ${\bf G}(10)<0.003$ for states marked with 
full circles, and ${\bf G}(10)>0.1$ for all other states (squares).

Note that the set of angular momentum multiplets which form subspace 
$\tilde{\bf H}_p$ of $N$ electrons each with angular momentum $l$ is 
always the same as the set of multiplets in subspace $\tilde{\bf 
H}_{p+1}$ of $N$ electrons each with angular momentum $l+(N-1)$.
When $l$ is increased by $N-1$, an additional band appears at high 
energy, but the structure of the low-energy part of the spectrum is 
completely unchanged.
For example, all three allowed multiplets for $l=5/2$ ($L=0$, 2, and 4)
form the lowest energy band for $l=11/2$, 17/2, and 23/2, where they 
span the $\tilde{\bf H}_1$, $\tilde{\bf H}_2$ and $\tilde{\bf H}_3$ 
subspace, respectively.
Similarly, the first excited band for $l=11/2$ is repeated for $l=17/2$ 
and 23/2, where it corresponds to $\tilde{\bf H}_1$ and $\tilde{\bf 
H}_2$ subspace.

Let us stress that the fact that identical sets of multiplets occur in 
subspace $\tilde{\bf H}_p$ for a given $l$ and in subspace $\tilde{\bf 
H}_{q+1}$ for $l$ replaced by $l+(N-1)$, does not depend on the form of 
interaction, and follows solely from the rules of addition of angular 
momenta of identical Fermions.
However, if the interaction pseudopotential has SRR, then:
(i) $\tilde{\bf H}_p$ are interaction eigensubspaces;
(ii) energy bands corresponding to $\tilde{\bf H}_p$ with higher $p$ 
lie below those of lower $p$;
(iii) spacing between neighboring bands is governed by a difference
between appropriate pseudopotential coefficients; and
(iv) wavefunctions and structure of energy levels within each band are
insensitive to the details of interaction.
Replacing $V_{\rm SRR}$ by a pseudopotential that increases more slowly
with increasing $L'$ leads to:
(v) coupling between subspaces $\tilde{\bf H}_p$;
(vi) mixing, overlap, or even order reversal of bands;
(vii) deviation of wavefunctions and the structure of energy levels 
within bands from those of the hard core repulsion (and thus their
dependence on details of the interaction pseudopotential).
The numerical calculations for the Coulomb pseudopotential in the lowest 
LL show (to a large extent) all SRR properties (i)--(iv), and virtually 
no effects (v)--(vii), characteristic of 'non-SRR' pseudopotentials.

The reoccurrence of $L$ multiplets forming the low-energy band when 
$l$ is replaced by $l\pm(N-1)$ has the following crucial implication.
In the lowest LL, the lowest energy ($p$th) band of the $N$ electron 
spectrum at the monopole strength $2S$ contains $L$ multiplets which 
are all the allowed $N$ electron multiplets at $2S-2p(N-1)$.
But $2S-2p(N-1)$ is just $2S^*$, the effective monopole strength of CF's!
The MFCS transformation which binds $2p$ fluxes (vortices) to each 
electron selects the same $L$ multiplets from the entire spectrum as 
does the introduction of a hard core, which forbids a pair of electrons 
to be in a state with $L'>2(l-p)$.

\section{Definition of short-range repulsive pseudopotential}

A useful operator identity relates total ($L$) and pair ($\hat{L}_{ij}$) 
angular momenta\cite{parent}
\begin{equation}
   \sum_{i<j} \hat{L}_{ij}^2 = \hat{L}^2 + N(N-2)\;\hat{l}^2.
\end{equation}
It implies that interaction given by a pseudopotential $V(L')$ that is 
linear in $\hat{L}'^2$ (e.g. the harmonic repulsion within each LL) is 
degenerate in each $L$ subspace and its energy is a linear function of 
$L(L+1)$.
The many-body GS has the lowest available $L$ while the maximum $L$ 
corresponds to the largest energy.
Note that this result is opposite to the Hund rule valid for spherical
harmonics, due to the opposite behavior of $V(L')$ for the FQH ($n=0$ 
and $l=S$) and atomic ($S=0$ and $l=n$) systems.

Deviations of $V(L')$ from a linear function of $L'(L'+1)$ lead to 
the level repulsion within each $L$ subspace, and the GS is no longer 
necessarily the state with minimum $L$.
Rather, it is the state at a low $L$ whose multiplicity $N_L$ (number of 
different $L$ multiplets) is large.
It interesting to observe that the $L$ subspaces with relatively high 
$N_L$ coincide with the MFCF prediction.
In particular, for a given $N$, they reoccur at the same $L$'s when 
$l$ is replaced by $l\pm(N-1)$, and the set of allowed $L$'s at 
a given $l$ is always a subset of the set at $l+(N-1)$.

As we said earlier, if $V(L')$ has short range, the lowest energy states 
within each $L$ subspace are those maximally avoiding large $L'$, and the 
lowest band (separated from higher states by a gap) contains states 
in which a number of largest values of $L'$ is avoided altogether.
This property is valid for all pseudopotentials which increase more
quickly than linearly as a function of $L'(L'+1)$.
For $V_\beta(L')=[L'(L'+1)]^\beta$, exponent $\beta>1$ defines the 
class of SRR pseudopotentials, to which the MFCF picture can be applied.
Within this class, the structure of low-lying energy spectrum and the
corresponding wavefunctions very weakly depend on $\beta$ and converge to 
those of $V_{\rm SRR}$ for $\beta\rightarrow\infty$.

The extension of the SRR definition to $V(L')$ that are not strictly in 
the form of $V_\beta(L')$ is straightforward.
If $V(L')>V(2l-m)$ for $L'>2l-m$ and $V(L')<V(2l-m)$ for $L'<2l-m$ and
$V(L')$ increases more quickly than linearly as a function of $L'(L'+1)$ 
in the vicinity of $L'=2l-m$, then pseudopotential $V(L')$ behaves like 
SRR at filling factors near $\nu=1/m$.

\section{Application to various interacting systems}

It follows from Fig.~\ref{fig2}a that the Coulomb pseudopotential in 
the lowest LL satisfies the SRR condition in the entire range of $L'$;
this is what validates the MFCF picture for filling factors $\nu\le1$.
However, in a higher, $n$th LL this is only true for $L'<2(l-n)-1$
(see Fig.~\ref{fig2}b for $n=1$) and the MFCF picture is valid only for 
$\nu_n$ (filling factor in the $n$th LL) around and below $(2n+3)^{-1}$.
Indeed, the MFCF features in the ten electron energy spectra around 
$\nu=1/3$ (in Fig.~\ref{fig1}) are absent for the same fillings of the 
$n=1$ LL.\cite{parent}

\begin{figure}[t]
\epsfxsize=3.5in
\centerline{\epsfbox{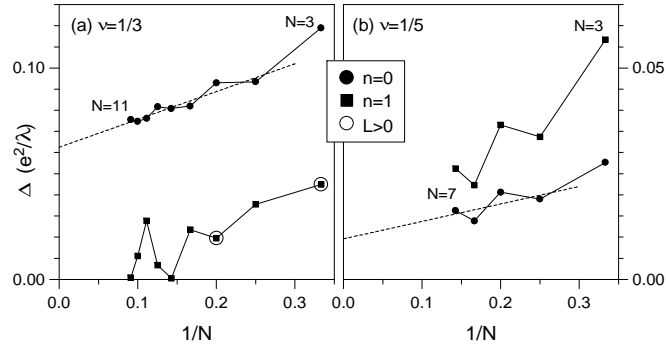}}
\caption{
   Excitation gap $\Delta$ as a function of inverse electron number 
   $1/N$ for filling factors $\nu=1/3$ (a) and 1/5 (b) in the $n=0$ 
   (dots) and $n=1$ (squares) LL's. 
   Open circles mark degenerate ground states ($L>0$).}
\label{fig4}
\end{figure}

One consequence of this is that the MFCF picture or Laughlin-like 
wavefunction cannot be used to describe the reported\cite{willet} 
incompressible state at $\nu=2+1/3=7/3$ ($\nu_1=1/3$).
The correlations in the $\nu=7/3$ GS are different than at $\nu=1/3$;
the origin of (apparent) incompressibility cannot be attributed to the 
formation of a Laughlin-like $\nu_1=1/3$ state on top of the 
$\nu=2$ state and connection between the excitation gap and the 
pseudopotential parameters is different.
This is clearly visible in the dependence of the excitation gap $\Delta$
on the electron number $N$, plotted in Fig.~\ref{fig4} for $\nu=1/3$ and 
1/5 fillings of the lowest and first excited LL.
The gaps for $\nu=1/5$ behave similarly as a function of $N$ in both 
LL's, while it is not even possible to make a conclusive statement
about degeneracy or incompressibility of the $\nu=7/3$ state based on
data for up to 11 electrons.

\begin{figure}[t]
\epsfxsize=3.5in
\centerline{\epsfbox{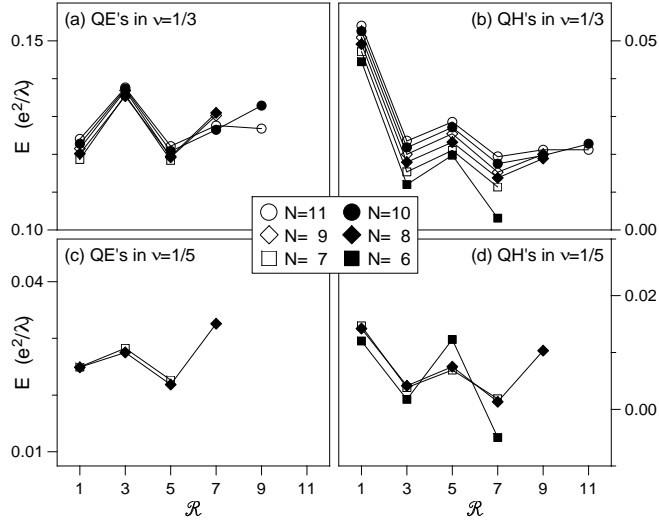}}
\caption{
   Energies of a pair of quasielectrons (left) and quasiholes (right) 
	in Laughlin $\nu=1/3$ (top) and $\nu=1/5$ (bottom) states, as a 
	function of relative pair angular momentum ${\bf R}$, obtained in 
	diagonalization of $N$ electrons.}
\label{fig5}
\end{figure}

The SRR criterion can be applied to the QP pseudopotentials to understand
why QP's do not form incompressible states at all Laughlin filling factors 
$\nu_{\rm QP}=1/m$ in the hierarchy picture\cite{haldane2,sitko} of FQH 
states.
Lines in Fig.~\ref{fig1}b and \ref{fig1}f mark $V_{\rm QE}$ and 
$V_{\rm QH}$ for the Laughlin $\nu=1/3$ state of ten electrons.
From similar calculations for diffent $N$ one can deduce the behavior 
of the QP pseudopotentials in the $N\rightarrow\infty$ limit.
Such analysis leads to a surprising conclusion that the SRR character
characteristic of the electron--electron pseudopotential in the lowest 
LL does not generally hold for the QP pseudopotentials.\cite{beran,yi}
In Fig.~\ref{fig5} we show the data for QE's and QH's in Laughlin 
$\nu=1/3$ (data for $N\le8$ was published before\cite{yi}) and 
$\nu=1/5$ states.
The plotted energy is given in units of $e^2/\lambda$ where $\lambda$ 
is the magnetic length in the parent state.
Different symbols mark pseudopotentials obtained in diagonalization 
of $N$ electron systems with different $N$ and thus with different 
$l_{\rm QP}$).
Clearly, the QE and QH pseudopotentials are quite different and neither
one decreases monotonically with increasing ${\bf R}$.
On the other hand, the corresponding pseudopotentials in $\nu=1/3$ 
and 1/5 states look similar, only the energy scale is different.
The convergence of energies at small ${\bf R}$ obtained for larger $N$ 
suggests that the maxima at ${\bf R}=3$ for QE's and at ${\bf R}=1$ 
and 5 for QH's, as well as the minima at ${\bf R}=1$ and 5 for QE's 
and at ${\bf R}=3$ and 7 for QH's, persist in the limit of large $N$ 
(i.e. for an infinite system on a plane).
Consequently, the only incompressible daughter states of Laughlin 
$\nu=1/3$ and 1/5 states are those with $\nu_{\rm QE}=1$ or $\nu_{\rm 
QH}=1/3$ (asterisks in Fig.~\ref{fig1}) and (maybe) $\nu_{\rm QE}=1/5$ 
and $\nu_{\rm QH}=1/7$ (question marks in Fig.~\ref{fig1}).
It is also clear that no incompressible daughter states will form at 
e.g. $\nu=4/11$ or 4/13.
Taking into account the behavior of involved QP pseudopotentials on 
all levels of hierarchy explains all observed odd-denominator FQH 
states and allows prediction of their relative stability (without 
using trial wavefunctions involving multiple LL's and projections 
onto the lowest LL needed in Jain's CF picture).

\section{Conclusion}

Using the pseudopotential formalism, we have described the FQH states 
in terms of the ability of electrons to avoid strongly repulsive pair 
states.
We have defined the class of SRR pseudopotentials leading to the formation 
of incompressible FQH states.
We argue that the MFCF picture is justified for the SRR interactions 
and fails for others.
The pseudopotentials of the Coulomb interaction in excited LL's and 
of Laughlin QP's in the $\nu=1/3$ state are shown to belong to the 
SRR class only at certain filling factors.

\section*{Acknowledgment}

AW and JJQ acknowledge partial support by the Materials Research
Program of Basic Energy Sciences, US Department of Energy.


\begin{thebibliography}{99}

\bibitem{klitzing} 
K. von Klitzing, G. Dorda, and M. Pepper,
   {\it Phys. Rev. Lett.} {\bf45}, 494 (1980).

\bibitem{tsui} 
D. C. Tsui, H. L. St\"ormer, A. C. Gossard,
   {\it Phys. Rev. Lett.} {\bf48}, 1559 (1982).

\bibitem{laughlin1} 
R. Laughlin, 
   {\it Phys. Rev. Lett.} {\bf50}, 1395 (1983).

\bibitem{prange} 
{\sl The Quantum Hall Effect}, 
edited by R. E. Prange and S. M. Girvin,
Springer-Verlag, New York (1987).

\bibitem{haldane1} 
F. D. M. Haldane and E. H. Rezayi,
   {\it Phys. Rev. Lett.} {\bf60}, 956 (1988).

\bibitem{parent}
A. W\'ojs and J. J. Quinn, 
   {\it Solid State Commun.} {\bf108}, 493 (1998);
   ibid. {\bf110}, 45 (1999);
   {\it Philos. Mag.} {\bf B80}, 1405 (2000);
J. J. Quinn, A. W\'ojs,
   {\it J. Phys.: Cond. Mat.} {\bf 12}, R265 (2000);
A. W\'ojs,
   {\it Phys. Rev.} {\bf B63}, 125312 (2001).
   
\bibitem{shalit}
A. de Shalit and I. Talmi, 
   {\sl Nuclear Shell Theory}, 
   Academic Press, New York (1963);
R. D. Cowan,
   {\sl The Theory of Atomic Structure and Spectra},
   University of California Press, 
   Berkeley (1981).

\bibitem{haldane2} 
F. D. M. Haldane, 
   {\it Phys. Rev. Lett.} {\bf51}, 605 (1983).

\bibitem{sitko}
P. Sitko, K.-S. Yi, and J. J. Quinn,
   {\it Phys. Rev.} B {\bf56}, 12417 (1997).

\bibitem{gasiorowicz}
S. Gasiorowicz, {\sl Quantum Physics}, 
John Wiley and Sons, New York (1974).

\bibitem{laughlin2}
R. B. Laughlin,
   {\it Surf. Sci.} {\bf142}, 163 (1984);
B. I. Halperin,
   {\it Phys. Rev. Lett.} {\bf52}, 1583 (1984);
J. K. Jain and V. J. Goldman,
   {\it Phys. Rev.} {\bf B45}, 1255 (1992).

\bibitem{willet}
R. Willet, J. P. Eisenstein, H. L. St\"ormer, D. C. Tsui, 
A. C. Gossard, and  J. H. English,
   {\it Phys. Rev. Lett.} {\bf59}, 1776 (1987).

\bibitem{mallet}
J. R. Mallet, R. G. Clark, R. J. Nicholas, R. L. Willet, 
J. J. Harris, and C. T. Foxon,
   {\it Phys. Rev.} {\bf B38}, 2200 (1988);
T. Sajoto, Y. W. Suen, L. W. Engel, M. B. Santos, and M. Shayegan,
   {\it Phys. Rev.} {\bf B41}, 8449 (1990).

\bibitem{jain}
J. Jain, 
   {\it Phys. Rev. Lett.} {\bf63}, 199 (1989).

\bibitem{lopez}
A. Lopez and E. Fradkin, 
   {\it Phys. Rev.} {\bf B44}, 5246 (1991).

\bibitem{wu}
T. T. Wu and C. N. Yang,
   {\it Nucl. Phys.} {\bf B107}, 365 (1976).

\bibitem{chen}
X. M. Chen and J. J. Quinn, 
   {\it Solid State Commun.} {\bf92}, 865 (1996).

\bibitem{beran} 
P. B\'eran and R. Morf,
   {\it Phys. Rev.} {\bf B43}, 12654 (1991).

\bibitem{yi}
S. N. Yi, X. M. Chen, and J. J. Quinn,
   {\it Phys. Rev.} {\bf B53}, 9599 (1996).

\end{thebibliography}
\end{document}